\documentstyle[12pt]{article}
\def\lsim{\mathrel{\rlap{\lower4pt\hbox{\hskip1pt$\sim$}}
    \raise1pt\hbox{$<$}}}         
\def\gsim{\mathrel{\rlap{\lower4pt\hbox{\hskip1pt$\sim$}}
    \raise1pt\hbox{$>$}}}         

\def\overleftrightarrow#1{\vbox{\ialign{##\crcr
    $\leftrightarrow$\crcr
    \noalign{\kern 1pt\nointerlineskip}
    $\hfil\displaystyle{#1}\hfil$\crcr}}}

\begin{document}
\hspace{12cm}{\bf NT@UW-99-37}\\

\begin{center}
{\bf Omega Meson Cloud and the Proton's Light Anti-Quark Distributions}
\end{center}
\begin{center}
{Mary Alberg$^{a,b}$, Ernest M. Henley$^{a,c}$, and Gerald A. Miller$^{a}$} 

{\small \em $^{a}$ Department of Physics, University of Washington, Seattle,
WA 98195, USA} \\
{\small \em$^b$ Department of Physics, Seattle University, Seattle,
WA 98122, USA } \\ 
{\small \em $^{c}$ Institute for Nuclear Theory, University of Washington, 
Seattle, WA 98195, USA }
\end{center}
\vspace{0.25 in}
\begin{abstract}

We use the meson cloud model of the nucleon
 to calculate distribution functions for $(\bar {d} - \bar{u})$ and $
\bar{d}/\bar{u}$ in the proton. Including the effect of the omega meson
cloud,  with a coupling constant $g_\omega^2/4\pi\approx 8$,
 allows  a reasonably good  description of the data.
\end{abstract}
\vspace {0.25 in}
There has been considerable interest in the flavor dependence of
the proton's  quark distributions. About ten years ago, 
the NMC measurement of the integral of
the $(\bar{d} - \bar{u})$ 
provided convincing  evidence for 
 the flavor asymmetry of the sea \cite{NMC}. The meson cloud model 
and the Sullivan process were used to explain the momentum fraction
distribution of $(\bar{d} - \bar{u})$. For  reviews, see
Refs.~\cite{S&T,sk}. 
Interest in the light sea quark distributions was heightened by the recent
measurement of the ratio $\bar{d}/\bar{u}$ by means of the Drell-Yan
process \cite {NA51,E866}. Similar results 
for $(\bar{d} - \bar{u})$ were obtained by HERMES
\cite{hermes}. In the context of
previous work on the light antiquark asymmetry in the nucleon, a good
description of the data  was
obtained for the $x$-dependence  (momentum fraction) of $(\bar{d} - \bar{u})$,
but not for the ratio of $\bar{d}/\bar{u}$ \cite{E866}. The meson cloud models 
provide ratios that
either increase monotonically with $x$ or turn back towards unity too slowly 
\cite {AH}. Various explanations have been advanced for the discrepancy, e.g.,
 effects of
the $\Delta$ \cite{Kumano:1991mj}-
\cite{MST},
and the influence of the Pauli exclusion principle \cite{MST},
adjustment of parameters \cite{NSSS}, but none of these changes provides a 
satisfactory  description of  the $\bar{d}/\bar{u}$ distribution function.
The use of a soft $\pi$N$\Delta$ form factor enables a reasonable description
of $(\bar{d}-\bar{u})$ \cite{E866}, but then $\bar{d}/\bar{u}$ is too large .
The importance of understanding the ratio $\bar{d}/\bar{u}$ was emphasized
recently \cite{awtpanic}.

In this Letter we show that the inclusion of 
the $\omega$ meson along with pion cloud effects,
provides a good description of  the present data for
both $(\bar{d} - \bar{u})$ and $\bar{d}/\bar{u}$.
The importance of the $\omega$ in providing the repulsive short-range
part of
the nucleon-nucleon and consequently attractive nucleon-antinucleon force 
has been known for a long time.
The Compton wavelength of this meson is
small,
but the large (almost puzzlingly large ${g_\omega^2\over 4\pi}\approx 20$)
coupling constants universally
used to describe the nucleon-nucleon scattering data cause the potential to
be non-vanishing over a fairly large range. The  $\omega$ meson also 
may   be very important in deep inelastic scattering from nuclei
\cite{gamlf}, so it is natural to consider its effects for a proton target.

We provide the usual formulae \cite{S&T}
for the effects of the meson cloud (generalized
Sullivan process) in order to facilitate
the reader's understanding of what follows. The wave function of the proton is
written in terms of Fock states with and without mesons
\begin{eqnarray}
\mid p\rangle = \sqrt{Z}\mid p\rangle_{\rm bare} + \sum_{MB}\int dy\;
d^2\vec{k}_\perp\; \phi_{BM}(y,k_\perp^2)
\mid  B(y,\vec{k}_\perp) M(1-y, - \vec{k}_\perp)\rangle \;.
\end {eqnarray}
Here Z is a wavefunction renormalization constant,  $\phi_{BM}(y,k_\perp^2) $
is the probability amplitude for finding a physical nucleon in a state
consisting of a baryon, $B$ with longitudinal momentum fraction y and meson
$M$ of 
momentum fraction (1-y) and squared transverse relative momentum $k_\perp^2$.

The quark distribution function $q(x)$ of a proton is given by
\begin{equation}
q(x) = q^{\rm bare}(x) + \delta q(x) \; ,
\end{equation}
with 
\begin {eqnarray}
\delta q(x) = \sum_{MB}
\left(\int_x^1 f_{MB}(y) q_M (\frac{x}{y})\frac{dy}{y} \; 
+\int_x^1 f_{BM}(y) q_B(\frac{x}{y})\frac{dy}{y}\right),
\end{eqnarray}
\begin{equation}
f_{MB}(y) = f_{BM}(1-y) \; ,\label{sym}
\end{equation}
and
\begin{equation}
f_{BM}(y) = \int_0^\infty\mid\phi_{BM}(y,k_\perp^2)\mid^2~d^2k_\perp\;.
\end{equation}
The cut-offs required in the model are taken from ref. \cite {S&T} 
\begin{equation}
G_M (t,u) = \exp{\frac {t- m_M^2}{2 \Lambda_M^2}}
\exp{{ u - m_B^2\over 2\Lambda_M^2}} \; ,
\end{equation}
where $\Lambda_M$ is a cut-off parameter for each meson and $t$ and $u$
are the usual kinematical
variables, expressed in terms of $\vec{ k}_\perp$ and $y$.
Such a form is required to respect the identity
(\ref{sym}) \cite{S&T, Szczurek:1996ur}.
The expressions for the
splitting functions  $f_{MB}(y)$ are those  given by \cite {S&T}
as derived in 
Ref.~\cite{holtmann}.
We include specifically $\pi$, $\omega$, and $\eta$ mesons, but the latter 
is negligible.
In the present paper, we  omit the effects of the  
$\rho$ meson as well as those of the intermediate  $\Delta$.
The former increases $(\bar{d} - \bar{u})$, whereas the
latter decreases it, so these effects tend to cancel. These effects have
been included by previous authors, and do not provide a satisfactory
description
of the ratio $\bar {d}/\bar {u}$.
The  number of each type of
meson,
$n_M$, is obtained by integrating the square of
$f_{MB}(y)$ over $y$.  Then for us
$Z=1- 3n_{\pi^0} -n_\omega$.

We need to discuss the functions $q_M(x)$ and $q_p(x)$ 
of our calculation.  Those for the
nucleon and pion are measured, but 
the quark distribution  functions of the $\omega$ meson are unknown.
The bag model suggests that the structure functions of the $\omega$, 
$\rho $ and $\pi$
mesons are the same. The near equality is more believable for the
$\omega$ and $\rho$, than for that between the vector and pseudoscalar mesons.
However, it has been traditional \cite{trad} to assume that 
the structure function of the $\rho$ and $\pi$ are the same.
Thus we use \cite{sutton}
\begin{eqnarray}
x q_v(x)=0.99x^{0.61}(1-x)^{1.02}\\  
xq_{\rm sea}(x)=0.2(1-x)^{5.0},
\end{eqnarray}
for the valence and sea quark distributions of both the pion and
omega mesons. The bare nucleon sea is parametrized \cite{holtmann} as 
\begin{eqnarray}
x\bar{Q}_{\rm bare}(x)=0.11(1-x)^{15.8}\\
\bar{Q}_{\rm bare}=u_{\rm sea}=\bar{u}_{\rm sea}=d_{\rm sea}=\bar{d}_{\rm sea}
\end{eqnarray}

The value of $\Lambda_\pi = (0.83\pm0.05) $ GeV is chosen
  to reproduce the range of allowed values of the integral  \cite{E866}
$ 
    D\equiv \int_0^1 dx\; (\bar{d}(x)-\bar{u}(x))=0.100\pm0.018 $
  using
  the
   sum rule of Henley \& Miller \cite{HM}: $D={2\over 3}n_\pi=2n_{\pi^0}$.
  The parameter $\Lambda_\omega$ is expected to be larger than
  $\Lambda_\pi$ \cite{bonn}, and we used the range: 1.3 GeV
  $<\Lambda_\omega<$ 1.8
  GeV.

The results of our calculations for $\bar{d}(x)-\bar{u}(x)$
are shown in Fig.~1, and
those for the ratio $\bar{d}(x)/\bar u(x)$ are shown in Fig.~2.
The $\pi$-nucleon coupling constant  is taken as ${g_\pi^2\over 4\pi}=13.6$,
the $\omega$-proton coupling constant is taken to be $\frac{g^2_\omega}{4 \pi}$
=8.1. In Fig.~1 the solid line is for $\Lambda_\pi = 0.83 $ GeV, for which 
$\int_0^1 (\bar{d}(x) - \bar{u}(x)) dx =1.0$.
The dashed lines are for $\Lambda_\pi = 0.78$ GeV and 
$\Lambda_\pi = 0.88$ GeV, the range of values constrained by the experimental error of $\pm 0.18$ in $D$. The $\omega $ meson, like any other 
isoscalar meson, has no effect here, so that this curve is that from the 
pion alone.
In Fig.~2 the solid curve shown is for $\Lambda_\omega = 1.5$  GeV. The dashed line shows the 
effect of leaving out the $\omega$ cloud contribution. 

We have examined 
the effect of varying both $\Lambda_\omega$ and $g_\omega$ in the range 
$1.3$ Gev $\leq \Lambda \leq 1.8$ GeV and 
$7<{g_\omega^2\over 4\pi}<20$. The larger values of $g_\omega$
are favored in fits to
nucleon-nucleon scattering data using one-boson exchange
potentials \cite{bonn} and the smaller ones from dispersion relation
descriptions of forward nucleon-nucleon scattering. 
The larger values of $g_\omega$ and $\Lambda$ 
tend to give too small values of the ratio $\bar {d}/\bar{u}$, while decreasing
$g_\omega$ or $\Lambda$ causes the maximum value of $\bar{d}/\bar{u}$ to be
too large and to appear at too high a value of $x$. 

  The effects of $\eta$ mesons are not included in the curves shown in
  the figures. These provide only a 1 to 2\% change in
  $\bar{d}/ \bar{u}$
  for phenomenologically viable coupling constants. Although we also examined
  the change of distribution function for the bare quarks suggested by 
\cite{AH}, the effect is sufficiently small when added to that of the $\omega$
meson that we do not show it. 
  
It is clear from the figures  
that a good description of the present data is provided by the inclusion of 
the $\omega$ meson with a reasonable coupling constant, and that therefore 
the meson cloud
picture can be successful. As mentioned above there are a number of
effects discussed in the literature, not included here, which   contribute to
the proton sea. In addition, one could include the effects of the $\sigma$
meson, which would also tend to suppress  $\bar{d}/ \bar{u}$.
Including these effects is likely to improve the description of the data or
modify the parameters describing the $\omega$-nucleon interaction.

The existence of better data would provide a severe test of the present model,
and the prospects of such seem imminent \cite{prelim}. But 
it is fair to conclude that the use of the $\omega$ along with the previously
suggested meson cloud effects does allow for a good description of the
present data.

We thank the USDOE and the USNSF for partial support of this work. We thank
R.J.~Holt for a useful discussion. One of us (EMH) thanks J. Speth for a useful
discussion and for his hospitality at the KFZ in Juelich.

\newpage

\section*{Figure Captions}

FIGURE 1. Comparison of our meson cloud model with data \cite{E866} for 
$(\bar{d} - \bar{u})$. The 
solid line is for  $\Lambda_\pi = 0.83 $ GeV, for which 
$D=\int_0^1 (\bar{d} - \bar{u}) dx =1.0$.
The dashed lines are for $\Lambda_\pi = 0.78$ GeV and 
$\Lambda_\pi = 0.88$ GeV, the range of values constrained by the experimental error of $\pm 0.18$ in $D$.

$$ $$
\noindent
FIGURE 2. Comparison of our meson cloud model with data \cite{E866} for $\bar{d}/ \bar{u}$. The 
solid line ($\Lambda_\pi = 0.83 $)
is for $\frac{g^2_\omega}{4 \pi}=8.1$ and $\Lambda_\omega = 1.5 $ GeV. 
The dashed line shows our result if the $\omega$ cloud contribution is omitted.
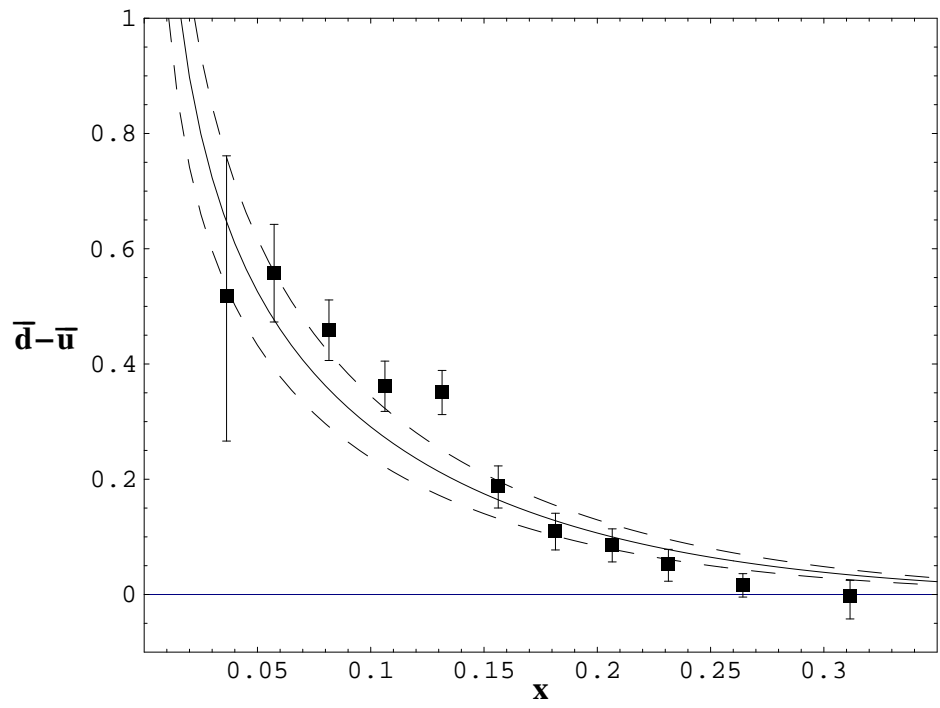
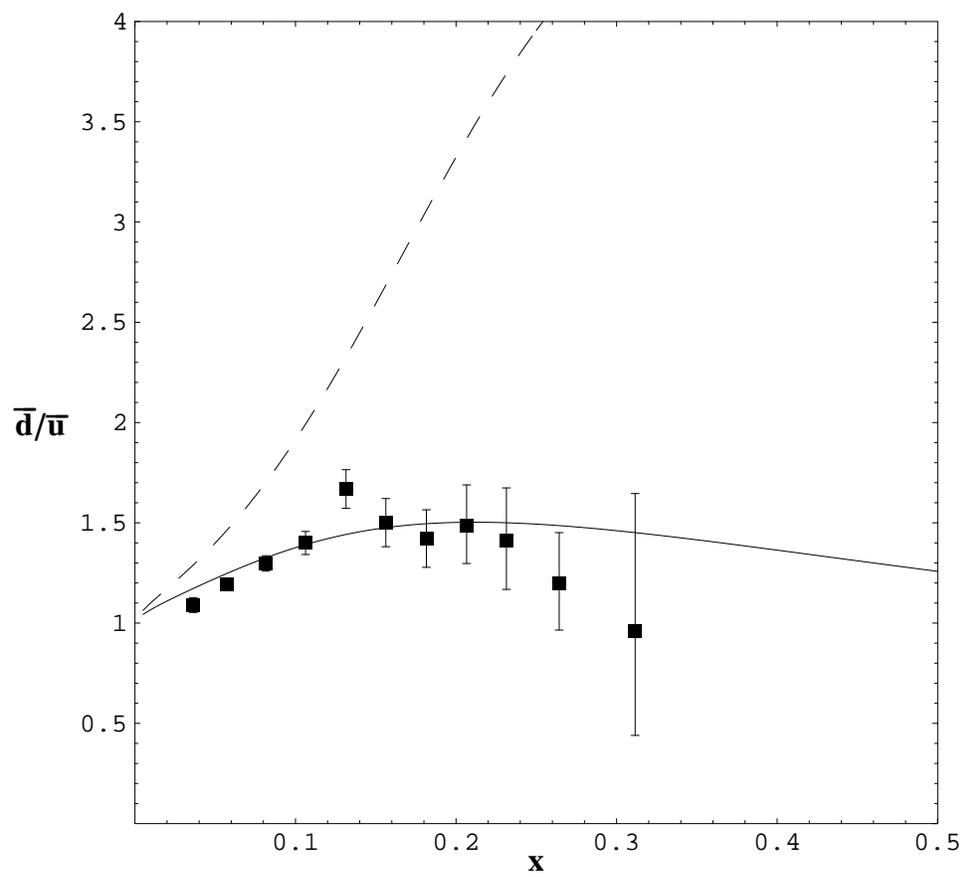
\end{document}